\documentclass[11pt]{article}
\usepackage[T1]{fontenc}
\usepackage{lmodern}
\usepackage[utf8]{inputenc}
\usepackage[margin=1in]{geometry}
\usepackage{amsmath,amssymb,graphicx,booktabs,url,listings}
\usepackage[colorlinks=true,allcolors=blue]{hyperref}
\title{On the use of meaningful score regions to interpret treatment effects on continuous clinical outcome assessments}
\author{
  Andrew Trigg\thanks{Clinical Statistics and Analytics, Bayer plc, Reading, UK. \href{mailto:andrew.trigg@bayer.com}{andrew.trigg@bayer.com}}
  \and
  Fraser D. Bocell\thanks{Clinical Outcome Assessment Program, Critical Path Institute, Tucson, AZ, USA.}
}
\date{}
\begin{document}
\maketitle
\begin{abstract}
Clinical outcome assessments (COAs), such as patient-reported outcomes (PROs), are faced with an interpretability issue due to a variety of score metrics being employed. Therefore, evidence linking COA scores to interpretable patient experiences is necessary to judge the extent to which the results of COA-based clinical trial endpoints are meaningful. The fourth Patient-Focused Drug Development guidance, published in draft form by the FDA in 2023, proposes the use of meaningful score regions (MSRs) to divide COA scores into more easily interpretable categories (e.g. severity-based MSRs of `none', `mild', `moderate' and `severe'). A recommendation is to compare the magnitude of estimated treatment effects for continuous COA scores (e.g. the difference in mean change from baseline between two treatment groups) to the maximum MSR width. In this article we explore this recommendation further, through historical evidence and simulations, and conclude that using the maximum MSR width as a benchmark may fail to detect meaningful treatment effects in practice. We suggest alternative approaches to evaluate treatment effects against MSRs, describing the expected outcome within each arm in terms of the probabilities of MSR membership.
\end{abstract}
\section{Introduction}
Many clinical outcome assessments (COAs), such as patient-reported outcomes (PROs), combine the responses of several items (i.e. questions or tasks) to yield scores on an unfamiliar metric which are difficult to interpret. While simple COAs based on a single item could produce readily interpretable scores, the meaning of a more complex COA score is not readily apparent in the same way as outcomes like survival or the occurrence of well-defined clinical events. The problem is exacerbated by the existence of multiple alternate PROs to measure the same concept (e.g. fatigue \cite{ref1}), each with different item content and scoring. Without guidelines to help with interpretation of the scores, it can be difficult to draw meaning from the scores. Given this interpretability issue, a large body of research has focused on interpretative thresholds for COAs~\cite{ref2,ref3,ref4}. The fourth and final guidance document arising from the FDA's Patient-Focused Drug Development (PFDD) initiative covers the topic of interpreting COA scores~\cite{ref5}. This draft guidance, released in April 2023 and subsequently referred to as `draft PFDD guidance \#4', splits the topic of COA interpretation into two broad approaches.

The first is referred to as the meaningful score difference (MSD) approach, which \emph{``identifies what size difference between any two COA scores would be viewed as meaningful for patients''}~\cite{ref5}. The MSD approach covers various types of interpretative thresholds, two of which are a meaningful within-patient change (MWPC) over time, and a meaningful between-group difference (MBGD) in change over time~\cite{ref4,ref6}. In a clinical trial setting, the MWPC is used to judge the extent to which an individual patient's change from baseline is meaningful and is commonly used to define responders.\footnote{Due to a loss of power associated with dichotomised responder analyses, it is recommended to reserve statistical testing for the continuous COA score and treat the responder analysis as supplementary information.} In a parallel-groups design, MBGD thresholds are used to judge the extent to which the difference in mean (change from baseline) scores between two treatment groups is meaningful. While draft PFDD guidance \#4 does not use the specific term MBGD, it cites one use of MSDs is \emph{``to evaluate the expected treatment effect for the average patient in some target population''}, \cite{ref5} which we consider MBGD to fall under. As such, MBGD thresholds enable the direct interpretation of continuous treatment effects estimated from a linear model (e.g. least squares [LS] mean difference from a mixed model for repeated measures). We do not distinguish between modelling the change from baseline score as the outcome, versus the score at each post-baseline visit, given both approaches yield an identical LS mean difference when baseline score is included as a covariate. A suitable MBGD will depend on the context of use including study population and whether marginal or conditional treatment effects are sought (given this choice can alter the population being targeted)~\cite{ref7}. Methods to estimate MWPC thresholds are well-developed and the subject of much research, \cite{ref2,ref3,ref4,ref8} in part due to the FDA's 2009 PRO guidance \cite{ref9} prioritising the use of MWPC to inform supplementary responder analyses. On the other hand, methods to estimate suitable MBGD thresholds are not yet established, where purely data-driven approaches are unlikely to be sufficient and subjective judgements are required~\cite{ref10}. The MBGD threshold for a given target population is likely to be smaller than the MWPC for the same target population \cite{ref10,ref11} and cannot realistically be any larger.\footnote{To illustrate this point, consider a MWPC of 5 points, and a hypothetical two-arm trial ($n=100$ per arm) where all patients receiving the experimental treatment have a score of 5 points ($=$MWPC) and those receiving control have a score of 0. We further assume, counterfactually, that those receiving experimental treatment would have scored 0 if assigned to control and those receiving control would have scored 5 if assigned to experimental treatment. This treatment effect (mean of 5 for treatment, mean of 0 for control, mean difference of 5) must be meaningful at the group-level as everybody on treatment versus nobody on control experiences a meaningful improvement; therefore, any MBGD threshold higher than the MWPC of 5 would overestimate the requirement. On the other hand, if ten patients on the experimental treatment had a score of 4 points instead, the estimated treatment effect (mean difference of 4.9) would still likely be understood as meaningful despite this falling below the MWPC of 5.} Therefore, the use of MWPC to directly interpret a continuous treatment effect is discouraged as it overestimates what is meaningful at the group-level. This issue of incorrectly applying MWPC thresholds to interpret a between-group difference has been described both in a general context \cite{ref10,ref11} and specific contexts of Alzheimer's disease \cite{ref12} and pain trials~\cite{ref13}.

\begin{figure}[htbp]
  \centering
  \includegraphics[width=0.92\linewidth]{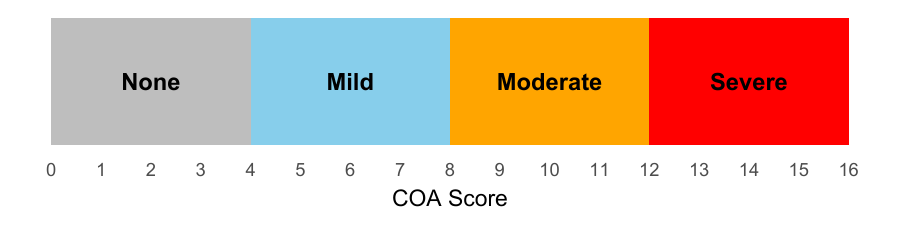}
  \caption{COA score partitioned into severity-based meaningful score regions.}
  \label{fig:one}
\end{figure}

The second broad approach to COA interpretation described in draft PFDD guidance \#4 is the meaningful score region (MSR) approach. The MSR approach aims to \emph{``specify the meaning of individual COA scores so that it is easier to judge whether two or more scores (e.g., treatment group means at a prespecified time point) correspond to distinct health-related experiences of patients''}~\cite{ref5}. While MSDs are based on interpreting differences between scores, MSRs aim to characterise the meaning of specific scores. Figure~\ref{fig:one} shows a COA score split into MSRs based on perceived severity. As such, MSRs are akin to severity cut-offs with which to categorise an individual's standing on the concept being measured. While the remainder of this paper refers to severity-based MSRs, we note that MSRs could also be based on aspects such as frequency, impairment, symptom control, or degree of bother; the issues discussed in this paper apply regardless.

The variety of approaches to interpret COA scores within draft PFDD guidance \#4 is a welcome expansion to the previous 2009 PRO guidance \cite{ref9} which focused on MWPC. We note that the guidance does not call for the ubiquitous application of both MSD and MSR approaches in future submissions, rather selection is based on an informed choice of which is most appropriate (or if both are appropriate).\footnote{The choice could be informed by aspects such as whether treatment is hypothesised to slow the decline of a condition (favouring MSD) versus improve it (where MSRs could additionally be useful), or whether the target population covers a limited versus broad range of severity (where utility of MSRs may be limited in a more restricted population).} In this article, we discuss specific recommendations on interpreting the extent to which a treatment effect is meaningful through the MSR approach. We focus on the mean difference in a COA score between two treatment arms, where the COA score is treated as continuous (such as a model-based LS mean difference) and refer to this as a continuous treatment effect throughout.

The remainder of this article is organised as follows. First, we highlight a recommendation in the draft PFDD guidance \#4 to compare continuous treatment effects to the width of MSRs. In the following two sections we assess the appropriateness of this recommendation; first, through examining historical treatment effects on established PRO measures; and second, through simulations. Based on this dual assessment, we conclude that the suggestion to compare continuous treatment effects to the MSR width imposes an unrealistically high hurdle. The subsequent section suggests alternate ways that MSRs can be applied to interpret COA scores within a parallel-groups randomised controlled trial, namely by focusing on the baseline-adjusted scores within each group rather than the difference between them.
\section{Comparing continuous treatment effects to MSR widths}
\begin{figure}[htbp]
  \centering
  \includegraphics[width=0.92\linewidth]{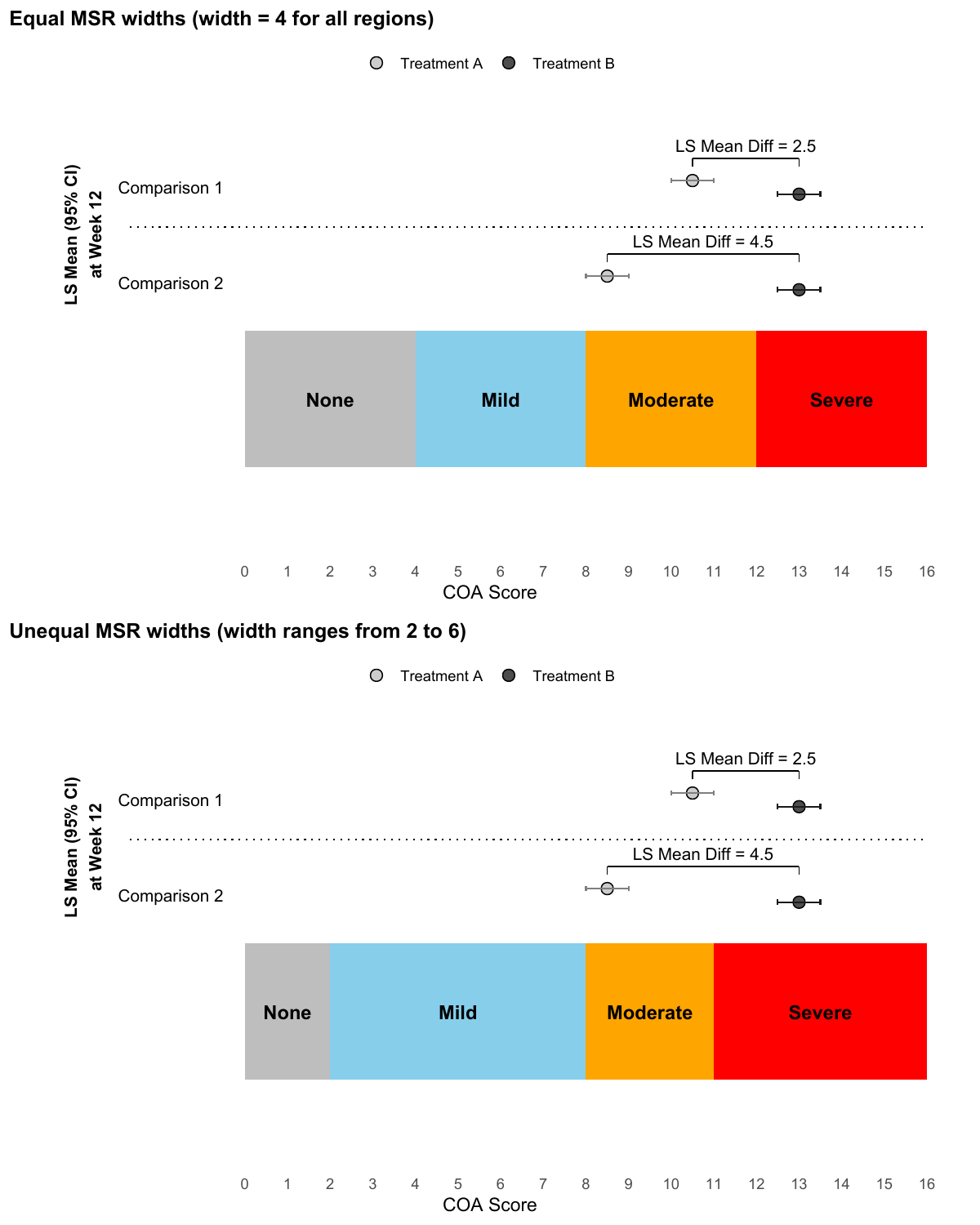}
  \caption{Continuous treatment effects compared to equal (top half) and unequal (bottom half) MSR widths. Treatment effects are compared with the maximum MSR width. Abbreviations: CI, confidence interval; COA, clinical outcome assessment; LS, least squares; MSR, meaningful score region.}
  \label{fig:two}
\end{figure}

Draft PFDD guidance \#4 suggests that continuous treatment effects (e.g. LS mean difference in change from baseline) can be compared to the widths of MSRs. If the MSR widths are approximately equal (top half of Figure~\ref{fig:two}), this common width is used for comparison. If the MSR widths are unequal (bottom half of Figure~\ref{fig:two}), the maximum MSR width is used, where \emph{``if the treatment effect is larger than the width of the widest score region, this suggests that the treatment effect reflects a meaningful difference to patients and/or caregivers''}~\cite{ref5}. We can continue to consider the maximum MSR width as a general case (as in the equal-width setting, all are at the maximum). Figure~\ref{fig:two} shows hypothetical continuous treatment effects mapped against MSRs. In Comparison 1, despite the COA scores of the average patient receiving Treatment A and B being in different regions, the difference of 2.5 points does not exceed the maximum MSR width (of 4 or 6 in the upper and lower plot respectively). In Comparison 2, the difference in average scores of 4.5 exceeds the maximum MSR width of 4 when the MSRs are equal, but fails to exceed the maximum MSR width of 6 (for the Mild region) when the MSRs are unequal.

\begin{figure}[htbp]
  \centering
  \includegraphics[width=0.92\linewidth]{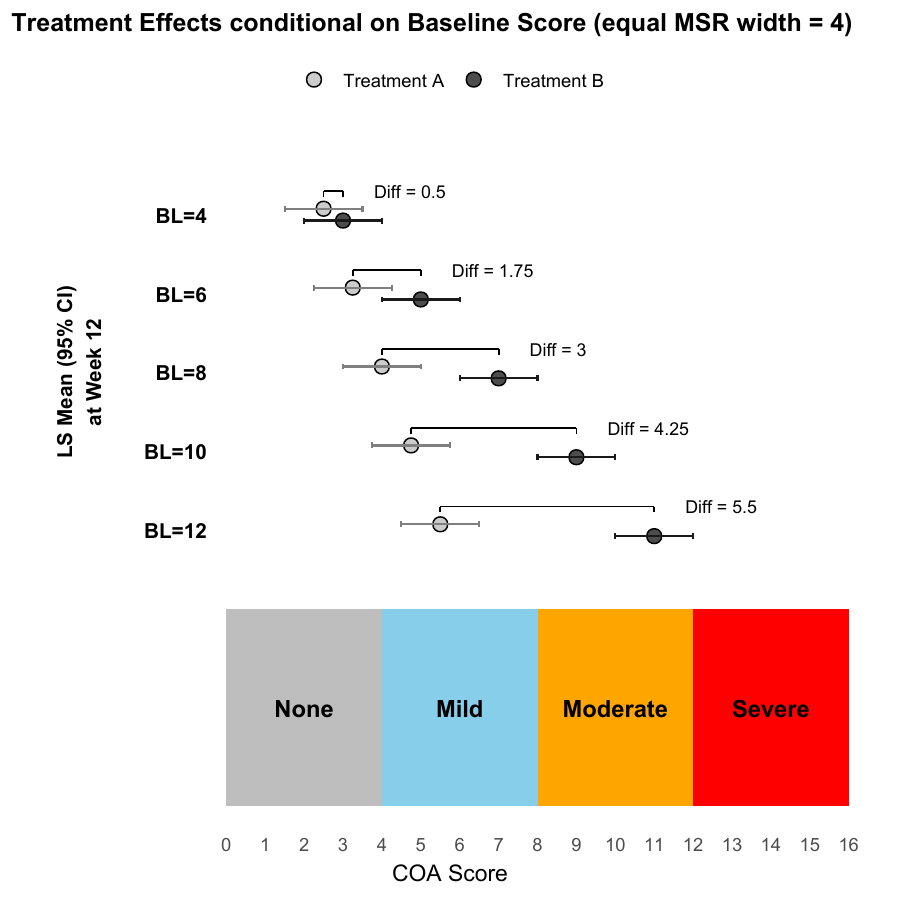}
  \caption{Continuous treatment effects, conditional on baseline score, compared to MSRs. Abbreviations: BL, baseline; CI, confidence interval; COA, clinical outcome assessment; LS, least squares; MSR, meaningful score region.}
  \label{fig:three}
\end{figure}

When the continuous treatment effect is smaller than the maximum MSR width, rather than simply classifying as `not meaningful', draft PFDD guidance \#4 suggests \emph{``additional analyses may be necessary to understand the nature of the treatment effect, such as exploring predicted COA scores at follow-up for each study arm over a range of baseline COA scores. This analysis may help identify which, if any, COA values at baseline are associated with a treatment effect that crosses two or more MSRs''}~\cite{ref5}. In other words, rather than estimating a marginal treatment effect for the full study population, a series of treatment effects conditional on baseline score can be estimated and compared to the maximum MSR width as per Figure~\ref{fig:three}. The conditional treatment effects can be estimated using a model incorporating an interaction term between baseline score and treatment. For the theoretical example in Figure~\ref{fig:three} we see that the estimated treatment effect surpasses the maximum MSR width of 4 for baseline COA scores of 10 points or greater. The implications of this scenario where the treatment effect is considered meaningful for a subset of patients, for example a decision on whether to include such results in a product label, are currently unclear. However, we consider this scenario to be less than ideal, given the shift away from the original marginal effect of interest. In cases where none of the conditional treatment effects exceed the maximum MSR width, although not explicitly mentioned draft PFDD guidance \#4, presumably a `not meaningful' judgement would be reached.

A proposed approach to estimate the thresholds between MSRs (and thus their width) is through anchor-based methods~\cite{ref14,ref15}. An anchor is an external criterion used to categorise health into meaningfully distinct and interpretable states. The probability of belonging to each anchor category can then be modelled conditional on the COA score, for example using a multinomial logit model~\cite{ref15}. A common anchor, featured as an example in draft PFDD guidance \#4, is the patient global impression of severity (PGI-S). The PGI-S is a single item asking patients to rate their perceived severity on the concept being measured by the COA score. For example, a COA measuring fatigue would be accompanied by a PGI-S asking patients to rate the severity of their fatigue symptoms. PGI-S measures commonly have 4 categories (`none', `mild', `moderate', `severe').

Given an anchor measure with $k$ categories and the possible range of a COA score, the maximum MSR width satisfies
\begin{equation}
  \max(\text{MSR width}) \geq \frac{\text{score range}}{k}.
  \label{eq:msr-width}
\end{equation}

When the MSR widths are equal, the full range of the COA score will be divided into $k$ regions of equal width. However, when the MSR widths are unequal, the maximum observed MSR width must exceed the score range divided by number of anchor categories.

Alternate approaches to anchor-based estimation of MSRs are cited in draft PFDD guidance \#4. The bookmarking method has patients and/or other stakeholders sort vignettes of patient experiences into ascending order of severity, then place `bookmarks' distinguishing regions (e.g. separating `Mild' from `Moderate')~\cite{ref16,ref17}. In this case, $k$ in the above formula is replaced by \emph{[number of bookmarks + 1]}. The use of an illustrative item from the COA of interest, to serve as a form of internal anchor, is also cited. In this case, $k$ in the above formula represents the number of categories for the illustrative item. The use of multiple illustrative items is also possible, where the maximum MSR width will vary by item depending on the number of categories for each.

\section{Comparing historical treatment effects to MSR widths}
In the previous section we established that the maximum MSR width must be equal or greater to the COA score range divided by the number of anchor categories (or equivalently the number of regions sought). We now put this into perspective with three case studies on well-established PRO measures, comparing known treatment effects to the MSR width.

The Kansas City Cardiomyopathy Questionnaire (KCCQ) is commonly used in heart failure and cardiomyopathy to assess heart failure symptoms and their impacts on functioning in patients with heart failure~\cite{ref18}. Three scores of the KCCQ have been qualified by the FDA for use in patients with heart failure~\cite{ref19}. All KCCQ scores range from 0 to 100. Three cardiomyopathy therapies with KCCQ treatment effects in the FDA product label are: tafamidis meglumine \cite{ref20} (treatment difference of 14 points on the Overall Summary score reported in the label), mavacamten \cite{ref21} (treatment difference of 9 points on the Clinical Summary Score reported in the label), and acoramidis \cite{ref22} (treatment difference of 10 points on the Overall Summary score reported in the label). If using a 4-category or 5-category PGI-S anchor, the maximum MSR width will be at least 25 or 20 points respectively, greatly exceeding the treatment differences reported in the product labels. However, the inclusion of KCCQ results in the product label presumably reflects their perceived meaningfulness, where smaller treatment effects not considered meaningful would not feature in the label.

A 0-10 itch numeric rating scale has been employed as a secondary endpoint in pivotal studies of FDA-approved systemic therapies for atopic dermatitis, such as dupilumab, tralokinumab, lebrikizumab, nemolizumab, abrocitinib, and upadacitinib~\cite{ref23,ref24,ref25,ref26,ref27,ref28}. A responder analysis based on the itch numeric rating scale ($\geq$4-point improvement from baseline) is described and presented in each product label of the above treatments. If instead using an MSR-based approach, using a 4-category PGI-S anchor, the maximum MSR width will be at least 2.5 points. However, a recent meta-analysis of systemic therapies has shown that only one FDA-approved therapy (upadacitinib in its highest dose) has an estimated treatment effect versus placebo exceeding 2.5 points~\cite{ref29}.

The European Organisation for Research and Treatment of Cancer Quality of Life Group Core 30 (EORTC QLQ-C30) is a widely used questionnaire of cancer symptoms and impacts, with all domain scores ranging from 0-100~\cite{ref30}. Therefore, using an anchor such as PGI-S with 4 categories yields a maximum MSR width of at least 25 points. A meta-analysis eliciting expert opinion on the magnitude of between-group mean differences in EORTC QLQ-C30 scores identified score differences corresponding to `trivial', `small', `medium' and `large' effects~\cite{ref31}. A 25-point difference exceeds the benchmarks for a `large' effect in all but one case (11 out of 12 scores for which benchmarks were fully estimable). If using a 5-category anchor and thus a maximum MSR width of at least 20 points, this also exceeds the `large' effect for most scores (8 out of 12).

Based on the above historical evidence, there are several cases where the continuous treatment effect would fail to be considered meaningful when setting a threshold based on MSR width, despite contrasting evidence supporting that these effects were meaningful to patients. In other words, the maximum MSR width may overestimate what is truly meaningful and its use as a benchmark for continuous treatment effects poses an unrealistically high goal to achieve.

We acknowledge, and draft PFDD guidance \#4 states, that the overall judgement whether a treatment provides meaningful benefit to patients involves careful consideration of multiple sources of information (i.e. other endpoints, applying multiple MSD or MSR thresholds, sensitivity analyses). However, some judgement is still required at the individual level of a COA-based endpoint, to provide the relevant piece of information to be considered among others. In a study with a COA-based primary endpoint, the extent to which the treatment effect on this endpoint is considered meaningful should be the predominant factor. We reiterate that failure to conclude that a treatment effect less than the maximum MSR width is meaningful does not necessarily result in the opposite conclusion of `not meaningful'; there is a middle-ground of uncertainty where effects conditional on baseline are explored.
\section{Simulations comparing MSR width and MWPC}
We supplement the above anecdotal evidence with simulations, comparing the maximum MSR width to a MWPC threshold estimated from the same anchor (e.g. PGI-S). Given the MWPC is reported to overestimate what is required to indicate a meaningful effect at the between-group level, if the MSR width is even larger we can be confident this approach overestimates what is required.
\subsection{Data generation}
Data were simulated for a PRO with 5 items, each of which has 4 categories (scored 0-3), administered at two visits. In addition, responses to a PGI-S with 4 categories (`None', `Mild', `Moderate', `Severe') were simulated at each visit. The process to simulate data was as follows, based on the assumption that a continuous latent trait variable exists, representing disease severity at each visit. The latent trait $\theta$ for each of 1000 individuals $i$ and each visit $q$ was simulated based on a bivariate normal distribution:

\begin{equation}
\begin{aligned}
 \boldsymbol{\theta}_{iq} &= [\theta_{i1},\theta_{i2}] \sim \operatorname{MVN}(\boldsymbol{\mu},\boldsymbol{\Sigma}),\\[4pt]
 \boldsymbol{\mu} &= \begin{bmatrix}0\\[2pt]0+\Delta\end{bmatrix},\\[4pt]
 \boldsymbol{\Sigma} &= \begin{bmatrix}1 & 0.5\\[2pt]0.5 & 1\end{bmatrix}.
\end{aligned}
\end{equation}
The parameter $\Delta$ denotes the extent of change from the first to second visit, which was varied in the simulations according to a uniform distribution between -1 and 0. Conditional on the latent trait $\theta$, responses to the 5 PRO items and additional PGI-S item were simulated based on a generalised partial credit model \cite{ref32}, an item response theory model where the probability of person $i$ providing a response in item $j$'s $k$th category, $y_{ijk}$ is:
\begin{equation}
 P(y_{ijk}\mid\theta_i,\alpha_j,\delta_{jk})=
 \frac{\exp\left[\sum_{h=0}^{k_j}\alpha_j(\theta_i-\delta_{jh})\right]}
 {\sum_{c=0}^{m_j}\exp\left[\sum_{h=0}^c\alpha_j(\theta_i-\delta_{jh})\right]}.
\end{equation}
In the above, $\alpha$ is the item discrimination parameter which governs the strength of relationship between the latent trait and probability of item response. Discrimination $\alpha$ was set to 1 for each of the PRO items, and allowed to vary for the PGI-S according to a uniform distribution between 0.5 and 1. Item $j$ is scored $0, 1, 2, \ldots, m_j$. The $h$th $\delta$  parameter is the location parameter between adjacent categories, representing the location on the latent trait where the probability of responding in adjacent categories is equal. As $\delta_{j0}$ is constrained to zero and each item has 4 categories, the location parameters $\delta_{j1}$, $\delta_{j2}$ and $\delta_{j3}$ are estimated. $c$ is an index used to sum across all possible categories in the denominator. The $\delta$ parameters were fixed across simulations for each PRO item, but varied for the PGI-S where $\delta_{j2}=0$, $\delta_{j1}=\delta_{j2}-\gamma$ and $\delta_{j3}=\delta_{j2}+\gamma$, and $\gamma$  follows a uniform distribution between 2 and 0.5. Varying the extent of change, PGI-S discrimination and PGI-S locations lead to expected variation in estimated MSRs and MWPCs~\cite{ref14,ref33}. 1000 datasets were simulated in this manner.
\subsection{Estimating MSR and MWPC in simulated datasets}
We estimated three MSR thresholds (yielding 4 regions) and a MWPC threshold for each simulated dataset. MSRs distinguishing the PGI-S categories of `none', `mild', `moderate' and `severe' were estimated, using the method of Terluin et al~\cite{ref14}. This estimation method fits an item response model to the PRO and PGI-S data and calculates the expected PRO score associated with the location parameters corresponding to the PGI-S responses. While Terluin and colleagues \cite{ref14} fit a graded response model where the location parameters denote the boundary between cumulative categories (e.g. $\delta_{j3}$ represents equal probability of none/mild/moderate versus severe), here we use the generalised partial credit model where the locations denote the boundary between adjacent categories. We feel the latter is more intuitive and reflects other methods for setting boundaries between regions such as bookmarking~\cite{ref17}.

MWPC thresholds were estimated using the method of Bjorner et al. \cite{ref33} fitting a longitudinal graded response model to the PRO items and dichotomised PGI-S change ($\geq$1-category improvement versus no change or worsening). The location parameter for the dichotomised PGI-S change informs the MWPC threshold on the latent trait metric, which is then converted to the observed PRO score metric. As the PGI-S change is dichotomised, using a graded response versus generalised partial credit model will not have a major impact.

The primary metric of interest was whether the maximum MSR width exceeded the MWPC threshold. We  acknowledge that comparing the maximum MSR width to MWPC should ideally have used a MBGD threshold as comparator, given the latter is more aligned with group-level continuous treatment effects. However, anchor-based estimation methods for MBGD are not established and thus we rely on the MWPC as a proxy, recognising that MWPC$\geq$MBGD~\cite{ref10}.
\subsection{Simulation results}
The maximum MSR width exceeded the MWPC in 100\% of the 1,000 simulated datasets. The maximum MSR width was 2.33 points (or 88.7\%) higher than the MWPC on average. The influence of the varying simulation parameters on maximum MSR width and MWPC is presented in the Supplementary Appendix Section~\ref{sec:supp-simulation-parameters}.

As the MWPC is already likely to overestimate what is required to indicate a meaningful continuous treatment effect, and the maximum MSR width consistently exceeds this, we can conclude that comparing continuous treatment effects to the maximum MSR width will result in false-negative conclusions on the meaningfulness of these effects. Combined with the anecdotal evidence, this suggests the maximum MSR width is too high a hurdle to overcome in practice.
\section{Alternate approaches to apply MSRs to interpret continuous COAs}
The presentation of MSRs alongside study results is valuable to enrich interpretation. However, the previous two sections suggest that directly comparing the magnitude of a continuous treatment effect to the maximum MSR width can fail to identify meaningful effects. Therefore, in this section we suggest alternative approaches to interpret COA scores in a parallel groups design against MSRs.

We recommend displaying the average outcome within each treatment arm (rather than the between-arm difference) relative to the MSRs. This ensures that both the treatment outcome and MSRs are on the same metric of COA score, rather than mixing scores with score differences. The adjusted mean outcome (e.g. LS mean adjusted for baseline score and other covariates) should be focused on, to ensure the numerical difference between these values aligns with the estimated LS mean difference used for statistical testing. This is similar to the example figure in draft PFDD guidance \#4 (similar to the top half of Figure~\ref{fig:two} in this paper), except we do not advocate for the additional presentation of the continuous treatment effect and its comparison to maximum MSR width.

\begin{figure}[tbp]
  \centering
  \includegraphics[width=0.92\linewidth]{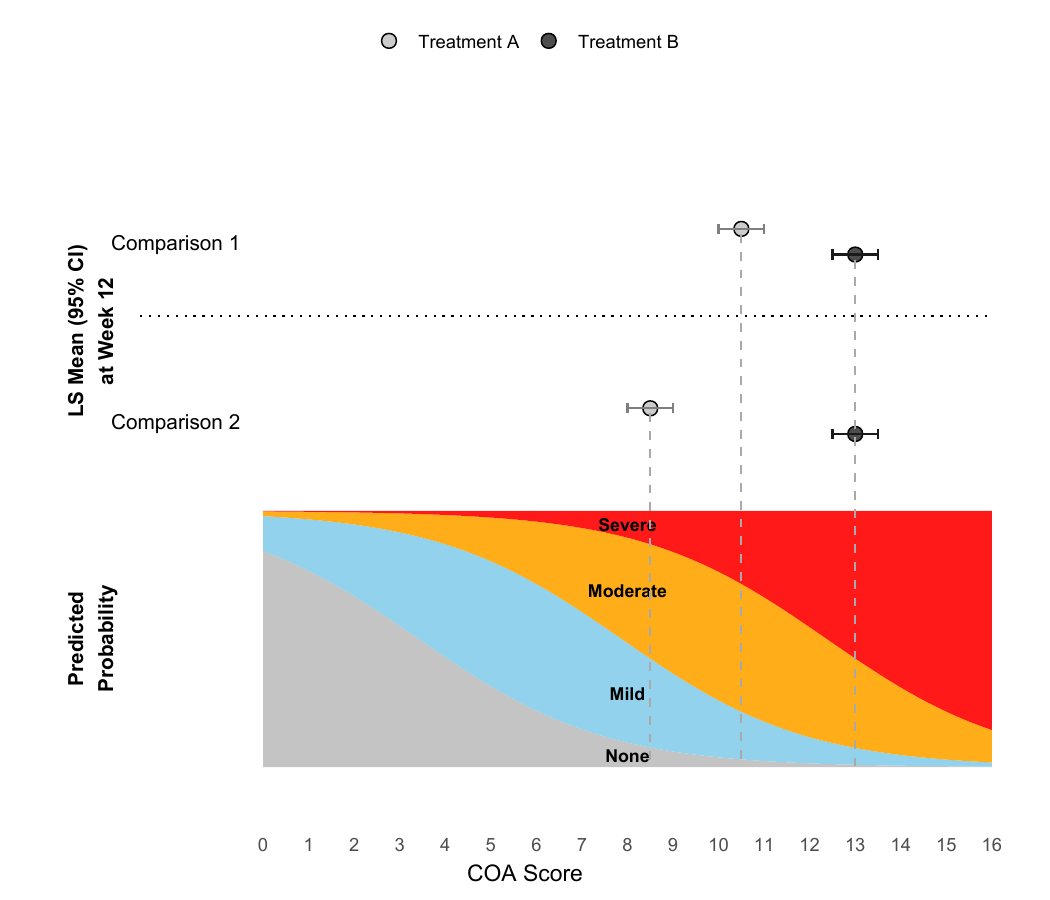}
  \caption{Average outcome for each treatment arm compared to predicted probability of region membership. Average outcomes are estimated by model-based LS means. Abbreviations: CI, confidence interval; COA, clinical outcome assessment; LS, least squares; MSR, meaningful score region.}
  \label{fig:four}
\end{figure}

\begin{figure}[tbp]
  \centering
  \includegraphics[width=0.92\linewidth]{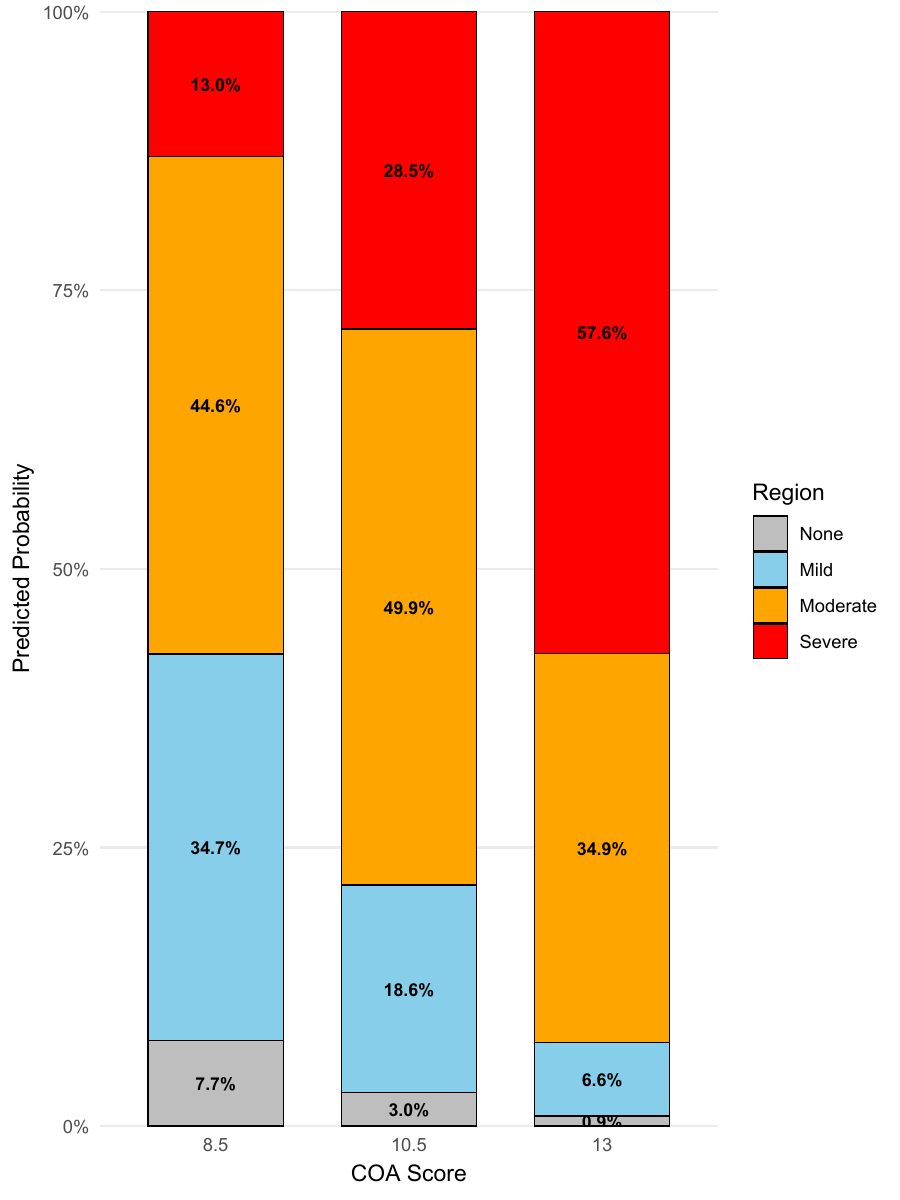}
  \caption{Probabilities of region membership for COA scores corresponding to average outcomes in each arm. Abbreviations: COA, clinical outcome assessment.}
  \label{fig:five}
\end{figure}

Once we focus on the (adjusted) mean outcome within each treatment arm, it is overly simplistic to state the most likely region associated with that score (e.g. in Figure~\ref{fig:one} to say a score of 9 is `Moderate' and a score of 7 is `Mild'). Therefore, we recommend presenting the probability of region membership given a particular score. Model-based predicted probabilities can be obtained when estimating the thresholds between MSRs~\cite{ref15}. Figure~\ref{fig:four} provides an example plot, similar to Figure~\ref{fig:two} but mapping the adjusted COA score for the average patient in each arm to the predicted probabilities of belonging in each region. In this manner, we see that the average scores for Treatment A (10.5 in Comparison 1 and 8.5 in Comparison 2) are no longer simply classified as `Moderate' but have distinct probabilities of region membership. Each possible score has a distinct interpretation rather than being categorised into the same region. An alternate display is shown in Figure~\ref{fig:five}, showing the predicted probabilities in a more easily readable format for the average scores of each treatment group in the previous example plots.

Notably, draft PFDD guidance \#4 suggests (in addition to directly interpreting the estimated treatment effect) plotting the empirical probability density function or cumulative distribution function of COA scores by treatment group, annotating with MSRs. Alternatively, the proportion of patients with post-baseline COA scores within specific regions can be presented by treatment arm, or the proportion with different category shifts from baseline. However, these approaches do not align with the estimand of primary interest subject to statistical testing.
\section{Discussion}
This article explores the use of MSRs to interpret continuous treatment effects on COA scores. The plausibility of comparing a treatment effect to the maximum MSR width was first considered in light of historical case studies, where each of these showed that treatment effects for approved drugs failed to meet this criterion. We acknowledge that these case studies are a subset of possible scenarios, and there may be instances with treatment effects exceeding the maximum MSR width.

Next, simulations were conducted to numerically compare the maximum MSR width to a MWPC threshold, both estimated using a PGI-S anchor. The logic behind this simulation was to uncover whether the MSR width is too high a requirement for a meaningful treatment effect, given the assumption that MWPC is already too high in many cases (given MWPC$\geq$MBGD). The MSR width was consistently higher than the MWPC, suggesting this is insensitive to assess the meaning of treatment effects. Many meaningful treatment effects would fail to be interpreted as meaningful when using this approach.

Given the above findings, we suggest moving away from comparing the treatment effect to maximum MSR width and instead focusing on the average scores within each arm. This ensures the same metric is employed for treatment outcomes and MSRs. It also uses adjusted mean outcomes aligned with the model used for statistical testing, rather than shifting away from the estimand of interest. By considering the probabilities of region membership for each possible score, rather than grouping several scores according to the single most likely region, we retain the full information along the score rather than categorising.

We acknowledge that the average outcome within each arm, and the thresholds between MSR regions, are both estimated with uncertainty; however, the simulations in Section 4 focus on comparing point estimates only. Future work should explore approaches to summarise and propagate this uncertainty.
\section{Conclusions}
MSRs represent a welcome framework providing more opportunity to evaluate the extent to which observed treatment effects on COA scores are meaningful. However, anecdotal and empirical evidence suggests continuous treatment effects (between-group differences in mean COA score) should not be compared to the maximum MSR width. Instead, the average outcome within each arm can be described in terms of the probabilities of MSR membership, enriching our understanding of how the average patient receiving each treatment feels or functions.

\section*{Acknowledgements}
An earlier draft of this paper benefitted greatly from the careful review and advice of Marc Vandemeulebroecke (Bayer), Kevin Weinfurt (Duke University) and an FDA reviewer. 

Figures 1 to 5 were produced in R using the ggplot2 package, where the R code was developed and refined with the assistance of Claude Opus 4.6. The code was reviewed and approved by the first author; the resulting figures were reviewed and approved by both authors.
\section*{Data availability statement}
No individual patient level data was analysed. Code to reproduce the simulations is provided in the Supplementary Appendix Section~\ref{sec:supp-r-code}.
\section*{Funding statement}
No external funding was received for this work.
\section*{Conflict of interest disclosure}
Andrew Trigg is an employee of Bayer plc and shareholder. Fraser Bocell is an employee of Critical Path Institute.

\clearpage
\appendix
\renewcommand{\thefigure}{\thesection\arabic{figure}}
\setcounter{figure}{0}
\section{Supplementary Appendix}
\subsection{Influence of varied simulation parameters on maximum MSR width and MWPC}
\label{sec:supp-simulation-parameters}
The influence of the varying parameters on maximum MSR width and MWPC across the 1,000 simulated datasets are presented below. While the PGI-S threshold parameter influences the MWPC in a linear fashion, the maximum MSR width reaches a minimum value for a given threshold when the MSRs are equally spaced, then increases either side of this.
\begin{figure}[htbp]\centering
\includegraphics[width=\linewidth]{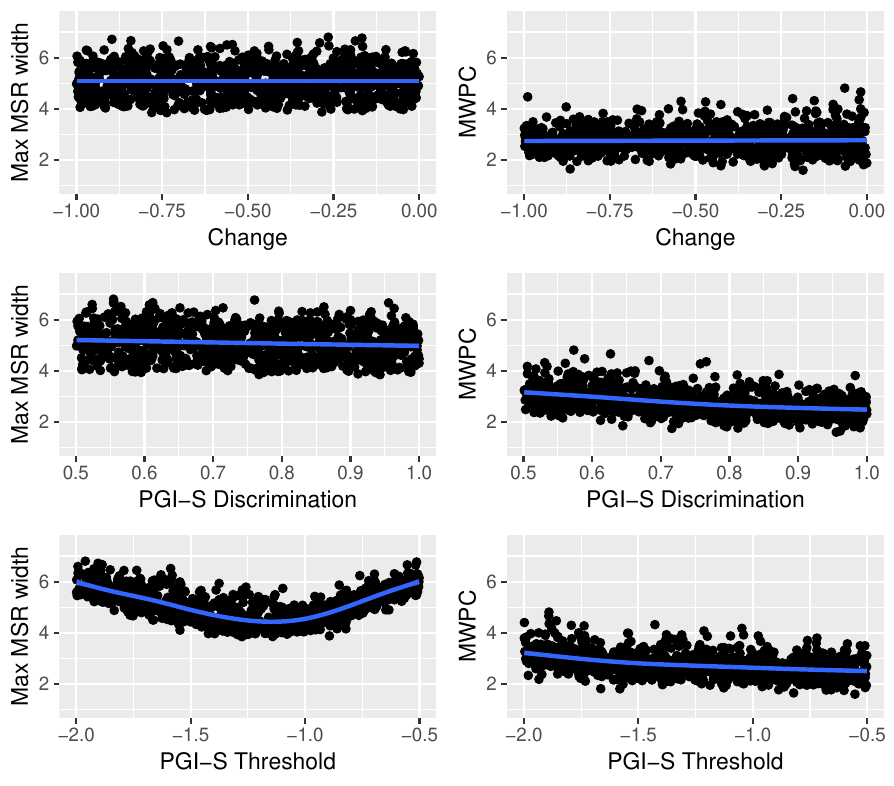}
\caption{Influence of varied simulation parameters on maximum MSR width and MWPC.}
\end{figure}
\clearpage
\subsection{R code for simulations}
\label{sec:supp-r-code}
Code to run the simulations described in Section 4 of the main text.
\begin{lstlisting}[language=R]
library(MASS)
library(polycor)
library(ggplot2)
library(tidyr)
library(dplyr)
library(mirt)
library(future)
library(future.apply)
library(gridExtra)

set.seed(20261)

simulatedata = function(chg, disc_pgi, thr_pgi){
dat=data.frame(mvrnorm(1000,
c(0,0+chg),
matrix(c(1,0.5,0.5, 1), nrow=2,byrow=T)))

dat$chg = dat$X2 - dat$X1

disc=c(1,1,1,1,1,disc_pgi) # 5 items with discrimination=1, then PGI with discrim=0.8
diff=matrix(c(0,2,3,3, # gpcm thresholds of -2, -1, 0
0,1.5,2,1.5, # gpcm thresholds of -1.5, -0.5, 0.5
0,1,1,0, # gpcm thresholds of -1, 0, 1
0,0.5, 0, -1.5, # gpcm thresholds of -0.5, 0.5, 1.5
0,0,-1,-3, # gpcm thresholds of 0, 1, 2
0,thr_pgi*-disc_pgi,thr_pgi*-disc_pgi,0 # thresholds for pgi (centered around 0)
),byrow=T,ncol=4)

# simulate item responses at Time 1 and Time 2
irtsim.T1 = simdata(a=disc,d=diff,Theta=dat$X1,itemtype='gpcm')
colnames(irtsim.T1) = c("i1.1", "i2.1", "i3.1", "i4.1","i5.1","pgi.1")
irtsim.T2 = simdata(a=disc,d=diff,Theta=dat$X2,itemtype='gpcm')
colnames(irtsim.T2) = c("i1.2", "i2.2", "i3.2", "i4.2","i5.2","pgi.2")

dat2=cbind(dat,irtsim.T1,irtsim.T2)

dat2$pgi.chg = dat2$pgi.2 - dat2$pgi.1 # PGI-S change
dat2$pgi.imp = ifelse(dat2$pgi.chg<=-1,1,0) # indicator for PGI-S improvement

corr_msr = polyserial(x=dat2$X1, y=dat2$pgi.1)
corr_mwpc = polyserial(x=dat2$chg, y=dat2$pgi.chg)

######### estimate MSRs ########

mod = mirt(dat2[,c(4:9)],itemtype="gpcm")

msr.latent = coef(mod,IRTpars=T)$pgi.1[2:4]
msr.score = expected.test(mod, Theta=coef(mod,IRTpars=T)$pgi.1[2:4],which.items=c(1:5))

width.latent = max(c(abs(msr.latent[2]-msr.latent[1]), abs(msr.latent[3]-msr.latent[2])))
width.score = max(c(
msr.score[1] - 0, # edge from min score=0 to first threshold
abs(msr.score[2]-msr.score[1]),  # in between 1 and 2
abs(msr.score[3]-msr.score[2]), # in between 2 and 3
15 - msr.score[3] # edge from 3rd threshold to max score=15
))

######### estimate MWPC ########

## Longitudinal IRT

dat3 = dat2[,c(4:8,10:14,17)]
nitems = 5
noptions=4
N.mic = 1000

prop_imp=mean(dat3$pgi.imp)

itemloadings = rep(1:5, times = 2)
itemloadings = c(itemloadings, NA)

model = 'Time1 = i1.1, i2.1, i3.1, i4.1, i5.1, pgi.imp
Time2 = i1.2, i2.2, i3.2, i4.2, i5.2, pgi.imp
COV = Time2*Time2, Time1*Time2
MEAN = Time2'

# construct constraints dynamically
# obtain starting values
sv = bfactor(dat3, itemloadings, model, pars='values')

# set up within time constraints
wtconstr = sv$parnum[(sv$name == 'a1' | sv$name == 'a2') & sv$est]

# create constraint list
constraints = list()
itemnames = colnames(dat3)
pick = c(0, nitems)

for(i in 1:nitems){
# accross time item constraints
constraints[[paste0('slope.', i)]] = sv$parnum[sv$name == paste0('a',2+i) & sv$est]
for(j in 1:(noptions-1)){
constraints[[paste0('intercept.', i, '_', j)]] =
sv$parnum[sv$name == paste0('d',j) & (sv$item %in% itemnames[pick + i]) & sv$est]
}
#across time constraints
constraints[[paste0('Time.', i)]] = wtconstr[pick + i]
}

( mod2 <- bfactor(dat3, itemloadings, model, constrain=constraints, TOL=5e-3,
itemtype = 'graded', optimizer = 'nlminb') )


## MIC in terms of theta change

cf = coef(mod2, simplify=TRUE)

( mwpc.latent = -cf$items[2*nitems+1,nitems+3]/cf$items[2*nitems+1,2] )

##### MIC in terms of scale score change

## Simulation method using the parameters of model

a = cf$items[,1:(nitems+2)]

d = cf$items[,(nitems+3):(nitems+5)]

th1 = rnorm(N.mic)
th2 = th1 + mwpc.latent
th3 = rnorm(N.mic)
th4 = rnorm(N.mic)
th5 = rnorm(N.mic)
th6 = rnorm(N.mic)
th7 = rnorm(N.mic)

Theta = data.frame(th1,th2,th3,th4,th5,th6,th7)
Theta = as.matrix(Theta)
dat1x = simdata(a,d,itemtype='graded',Theta=Theta)
dat1x = as.data.frame(dat1x)

sum1 = rowSums(dat1x[,1:nitems])

sum2 = rowSums(dat1x[,(nitems+1):(2*nitems)])

sumdif = sum2 - sum1

( mwpc.score.median =
median(sumdif) )  # MIC based on medium difference score
( mwpc.score.mean =
mean(sumdif) )    # MIC based on mean difference score

data.frame(
chg = chg,
disc_pgi = disc_pgi,
thr_pgi = thr_pgi,
msr.score1 = msr.score[1],
msr.score2 = msr.score[2],
msr.score3 = msr.score[3],
width.score = width.score,
mwpc.score.median = mwpc.score.median,
mwpc.score.median.abs = abs(mwpc.score.median),
mwpc.score.mean = mwpc.score.mean,
mwpc.score.mean.abs = abs(mwpc.score.mean),
widthbigger.score.median = width.score>abs(mwpc.score.median),
widthbigger.score.mean = width.score>abs(mwpc.score.mean),
diff.score.median = width.score - abs(mwpc.score.median),
diff.score.mean = width.score - abs(mwpc.score.mean),
corr_msr = corr_msr,
corr_mwpc = corr_mwpc,
prop_imp = prop_imp
)
}

###
# run simulations
###

set.seed(20261)

# Set up the plan for parallel processing
plan(multisession, workers=7)

# Run simulations in parallel
start = Sys.time()

out= list()

# Use future
for(i in 1:1000){
chg = runif(1,min=-1, max=0)
disc_pgi = runif(1, min=0.5, max=1)
thr_pgi = runif(1, min=-2, max = -0.5)
out[[i]] <- future({
simulatedata(chg=chg, disc_pgi=disc_pgi, thr_pgi=thr_pgi)
}, seed=TRUE)
}
out2=value(out)

end = Sys.time()
end-start

out3 = bind_rows(out2, .id="sim")


###
## summarise results
###


out4 =  out3 %>%
mutate(pct.diff.score = diff.score.mean/abs(mwpc.score.mean)) %>%
select(mwpc.score.mean,width.score,diff.score.mean,pct.diff.score)
summary(out4)

summary_results = out3 %>%
summarise(prop_widthbigger.score = mean(widthbigger.score.mean),
avg_diff.score = mean(diff.score.mean),
avg_pct.diff.score = mean(diff.score.mean/abs(mwpc.score.mean))
)

# plot estimates by varied parameters
width.chg = ggplot(out3, aes(x = chg, y = width.score)) +
geom_point() +
geom_smooth() +
scale_y_continuous(limits = c(1, 7.5)) +
labs(x = "Change", y = "Max MSR width")

mwpc.chg = ggplot(out3, aes(x = chg, y = mwpc.score.mean.abs)) +
geom_point() +
geom_smooth() +
scale_y_continuous(limits = c(1, 7.5)) +
labs(x = "Change", y = "MWPC")

width.disc = ggplot(out3, aes(x = disc_pgi, y = width.score)) +
geom_point() +
geom_smooth() +
scale_y_continuous(limits = c(1, 7.5)) +
labs(x = "PGI-S Discrimination", y = "Max MSR width")

mwpc.disc = ggplot(out3, aes(x = disc_pgi, y = mwpc.score.mean.abs)) +
geom_point() +
geom_smooth() +
scale_y_continuous(limits = c(1, 7.5)) +
labs(x = "PGI-S Discrimination", y = "MWPC")

width.thr = ggplot(out3, aes(x = thr_pgi, y = width.score)) +
geom_point() +
geom_smooth() +
scale_y_continuous(limits = c(1, 7.5)) +
labs(x = "PGI-S Threshold", y = "Max MSR width")

mwpc.thr = ggplot(out3, aes(x = thr_pgi, y = mwpc.score.mean.abs)) +
geom_point() +
geom_smooth() +
scale_y_continuous(limits = c(1, 7.5)) +
labs(x = "PGI-S Threshold", y = "MWPC")

# Combine plots
grid.arrange(
width.chg, mwpc.chg,
width.disc, mwpc.disc,
width.thr, mwpc.thr,
ncol = 2, nrow = 3
)

\end{lstlisting}

\begin{thebibliography}{99}
\bibitem{ref1} Hughes A, Ju A, Cazzolli R, et al. Patient-reported outcome measures for fatigue in patients with chronic kidney disease: a systematic review. BMJ Open. 2025;15(7):e099592. doi:10.1136/bmjopen-2025-099592
\bibitem{ref2} King MT. A point of minimal important difference (MID): a critique of terminology and methods. Expert Review of Pharmacoeconomics \& Outcomes Research. 2011;11(2):171-184. doi:10.1586/erp.11.9
\bibitem{ref3} Coon CD, Cook KF. Moving from significance to real-world meaning: methods for interpreting change in clinical outcome assessment scores. Qual Life Res. 2018;27(1):33-40. doi:10.1007/s11136-017-1616-3
\bibitem{ref4} Trigg A, Lenderking WR, Boehnke JR. Introduction to the special section: ``Methodologies and considerations for meaningful change.'' Qual Life Res. 2023;32(5):1223-1230. doi:10.1007/s11136-023-03413-1
\bibitem{ref5} FDA. Patient-Focused Drug Development: Incorporating Clinical Outcome Assessments Into Endpoints For Regulatory Decision-Making. Draft Guidance. Published online 2023. Accessed May 14, 2026. \url{https://www.fda.gov/media/166830/download}
\bibitem{ref6} Amdal CD, Falk RS, Alanya A, et al. SISAQOL-IMI consensus-based guidelines to design, analyse, interpret, and present patient-reported outcomes in cancer clinical trials. The Lancet Oncology. 2025;26(12):e683-e693. doi:10.1016/S1470-2045(25)00520-0
\bibitem{ref7} Weinfurt K. Interpreting the meaningfulness of treatment effects estimated in parallel groups designs: comment on Trigg et al. Qual Life Res. 2025;34(7):1885-1889. doi:10.1007/s11136-025-03952-9
\bibitem{ref8} Staunton H, Willgoss T, Nelsen L, et al. An overview of using qualitative techniques to explore and define estimates of clinically important change on clinical outcome assessments. Journal of Patient-Reported Outcomes. 2019;3(1):16. doi:10.1186/s41687-019-0100-y
\bibitem{ref9} FDA. Guidance for Industry Patient-Reported Outcome Measures: Use in Medical Product Development to Support Labeling Claims. Published online 2009. Accessed May 14, 2026. \url{https://www.fda.gov/media/77832/download}
\bibitem{ref10} Trigg A, Ayasse ND, Coon CD. Conceptualizing meaningful between-group difference in change over time: a demonstration of possible viewpoints. Qual Life Res. 2025;34(1):151-160. doi:10.1007/s11136-024-03798-7
\bibitem{ref11} McLeod LD, Cappelleri JC, Hays RD. Best (but oft-forgotten) practices: expressing and interpreting associations and effect sizes in clinical outcome assessments. The American Journal of Clinical Nutrition. 2016;103(3):685-693. doi:10.3945/ajcn.115.120378
\bibitem{ref12} Lansdall CJ, Cummings JL, Andrews JS. Appropriate use of meaningful within-patient change (MWPC) thresholds in Alzheimer's disease. Alzheimers Dement. 2025;21(2):e14436. doi:10.1002/alz.14436
\bibitem{ref13} Smith SM, Dworkin RH, Turk DC, et al. Interpretation of chronic pain clinical trial outcomes: IMMPACT recommended considerations. Pain. 2020;161(11):2446-2461. doi:10.1097/j.pain.0000000000001952
\bibitem{ref14} Terluin B, Koopman JE, Hoogendam L, Griffiths P, Terwee CB, Bjorner JB. Estimating meaningful thresholds for multi-item questionnaires using item response theory. Qual Life Res. 2023;32(6):1819-1830. doi:10.1007/s11136-023-03355-8
\bibitem{ref15} Yuan J, Lin L, Weinfurt K, et al. Meaningful Score Differences and Meaningful Score Regions of the Patient-Reported Outcomes Measurement Information System\textregistered{} Pediatric Asthma Impact Scale. Value Health. 2025;28(10):1540-1547. doi:10.1016/j.jval.2025.05.010
\bibitem{ref16} Morgan EM, Mara CA, Huang B, et al. Establishing clinical meaning and defining important differences for Patient-Reported Outcomes Measurement Information System (PROMIS\textregistered{}) measures in juvenile idiopathic arthritis using standard setting with patients, parents, and providers. Qual Life Res. 2017;26(3):565-586. doi:10.1007/s11136-016-1468-2
\bibitem{ref17} Cook KF, Cella D, Reeve BB. PRO-Bookmarking to Estimate Clinical Thresholds for Patient-reported Symptoms and Function. Med Care. 2019;57 Suppl 5 Suppl 1:S13-S17. doi:10.1097/MLR.0000000000001087
\bibitem{ref18} Green CP, Porter CB, Bresnahan DR, Spertus JA. Development and evaluation of the Kansas City Cardiomyopathy Questionnaire: a new health status measure for heart failure. J Am Coll Cardiol. 2000;35(5):1245-1255. doi:10.1016/s0735-1097(00)00531-3
\bibitem{ref19} FDA. Qualification of the Kansas City Cardiomyopathy Questionnaire Clinical Summary Score and its Component Scores. A Patient-Reported Outcome Instrument for Use in Clinical Investigations in Heart Failure. Published online 2020. Accessed May 14, 2026. \url{https://www.fda.gov/media/136862/download?attachment}
\bibitem{ref20} FDA. Tafamidis meglumine. Package Insert. Published online 2023. Accessed May 14, 2026. \url{https://www.accessdata.fda.gov/drugsatfda\_docs/label/2023/211996s002,212161s002lbl.pdf}
\bibitem{ref21} FDA. Mavacamten. Package Insert. Published online 2025. Accessed May 14, 2026. \url{https://www.accessdata.fda.gov/drugsatfda\_docs/label/2025/214998s010lbl.pdf}
\bibitem{ref22} FDA. Acoramidis. Package insert. Published online 2024. Accessed May 14, 2026. \url{https://www.accessdata.fda.gov/drugsatfda\_docs/label/2024/216540s000lbl.pdf}
\bibitem{ref23} FDA. Dupilumab. Package insert. Published online 2026. Accessed May 14, 2026. \url{https://www.accessdata.fda.gov/drugsatfda\_docs/label/2026/761055s074lbl.pdf}
\bibitem{ref24} FDA. Tralokinumab. Package Insert. Published online 2025. Accessed May 14, 2026. \url{https://www.accessdata.fda.gov/drugsatfda\_docs/label/2025/761180s019lbl.pdf}
\bibitem{ref25} FDA. Lebrikizumab. Package Insert. Published online 2025. Accessed May 14, 2026. \url{https://www.accessdata.fda.gov/drugsatfda\_docs/label/2025/761306s005lbl.pdf}
\bibitem{ref26} FDA. Nemolizumab. Package Insert. Published online 2025. Accessed May 14, 2026. \url{https://www.accessdata.fda.gov/drugsatfda\_docs/label/2025/761390s002,761391s001lbl.pdf}
\bibitem{ref27} FDA. Abrocitinib. Package Insert. Published online 2023. Accessed May 14, 2026. \url{https://www.accessdata.fda.gov/drugsatfda\_docs/label/2023/213871s004lblcorrected.pdf}
\bibitem{ref28} FDA. Upadacitinib. Package Insert. Published online 2025. Accessed May 14, 2026. \url{https://www.accessdata.fda.gov/drugsatfda\_docs/label/2025/211675Orig1s028,218347Orig1s005lbl.pdf}
\bibitem{ref29} Chu AWL, Wong MM, Rayner DG, et al. Systemic treatments for atopic dermatitis (eczema): Systematic review and network meta-analysis of randomized trials. Journal of Allergy and Clinical Immunology. 2023;152(6):1470-1492. doi:10.1016/j.jaci.2023.08.029
\bibitem{ref30} Aaronson NK, Ahmedzai S, Bergman B, et al. The European Organization for Research and Treatment of Cancer QLQ-C30: a quality-of-life instrument for use in international clinical trials in oncology. J Natl Cancer Inst. 1993;85(5):365-376. doi:10.1093/jnci/85.5.365
\bibitem{ref31} Cocks K, King MT, Velikova G, Martyn St-James M, Fayers PM, Brown JM. Evidence-based guidelines for determination of sample size and interpretation of the European Organisation for the Research and Treatment of Cancer Quality of Life Questionnaire Core 30. J Clin Oncol. 2011;29(1):89-96. doi:10.1200/JCO.2010.28.0107
\bibitem{ref32} de Ayala RJ. The Theory and Practice of Item Response Theory. Guilford Press; 2009.
\bibitem{ref33} Bjorner JB, Terluin B, Trigg A, Hu J, Brady KJS, Griffiths P. Establishing thresholds for meaningful within-individual change using longitudinal item response theory. Qual Life Res. 2023;32(5):1267-1276. doi:10.1007/s11136-022-03172-5
\end{thebibliography}
\end{document}